# Spectral emissivity of copper and nickel in the mid-infrared range between 250 and 900 °C

I. Setién-Fernández [a,b], T. Echániz [a,b], L. González-Fernández [c], R.B. Pérez-Sáez [a,b,⇑], M.J. Tello [a,b,d]

[a] Departamento de Física de la Materia Condensada, Facultad de Ciencia y Tecnología, Universidad del País Vasco, Barrio Sarriena s/n, 48940 Leioa, Bizkaia, Spain

[b] Instituto de Síntesis y Estudio de Materiales, Universidad del País Vasco, Apdo. 644, 48080 Bilbao, Spain

[c] CNRS, UPR3079 Conditions Extrêmes et Matériaux: Haute Température et irradiation, 1D Avenue de la Recherche Scientifique, 45071 Orléans Cedex 2, France

[d] CIC Energigune, Parque Tecnológico, Albert Einstein 48, 01510 Miñano, Álava, Spain

Abstract

A study on the radiative properties of two pure metals, copper and nickel, using a high accuracy radiometer is carried out. Their spectral emissivity between 3 and 21 μm and its dependence on emission angle and temperature between 250 and 900 °C is measured. An evolution of the samples emissivity associated to the surface stress relaxation is observed, which is relieved after two or three heating cycles. Spectral emissivity of metals usually decreases as wavelength increases, but in the case of copper an irregular behavior has been found. Its spectral emissivity shows a broad plateau around 10 μm, which can be due to the anomalous skin effect. On the other hand, the emissivity usually increases with temperature, but in the case of nickel the emissivity changes little and even slightly decreases for T > 700 °C. The experimental directional emissivity of both metals shows the dependence on the emission angle predicted by the electromagnetic theory for metallic samples. By increasing the emission angle, the emissivity dependence on the wavelength strongly decreases. Furthermore, in the case of nickel, an emissivity increase with wavelength is observed for λ > 20 μm. The electrical resistivity for both metals is obtained by fitting the experimental emissivity curves with the Hagen–Rubens equation. The results agree fairly well with direct electrical resistivity measurements for copper but show a poor agreement in the case of nickel.

1. Introduction

Knowledge of the radiative properties is essential for many applications related to the heat transfer in energy, industrial and scientific processes. Direct absolute measurements of the radiative properties are of great importance because of the theoretical relations between these physical quantities and the optical and electrical properties of materials [1,2]. In particular, experimental radiative measurements in metals have great interest for mid and far infrared frequencies as the direct measurement of optical constants on metals in this spectral range offers serious difficulties. Only room temperature experimental values of optical constants can be found [3–6]. This explains the discrepancies between the emissivity calculated by using experimental optic constants via the classical electromagnetic equations [1,2] and those experimental values obtained with a radiometer. On the other hand, some discrepancies appear because real materials do not meet the ideal surface conditions assumed by the theory. Their radiative and optical properties are highly influenced by the surface state, and depend on surface roughness, oxidation, surface contamination, impurities, thermal uniformity, etc. This fact explains the differences among experimental emissivity values found in the literature when the measurements are carried out with samples without a good characterization of their surface state or without adequate atmosphere control. At the present time, radiometers with a proper sample holder and a FT-IR spectrometer allow faster and more accurate measurements of the radiative properties as a function of wavelength, temperature and emission angle in a con- trolled atmosphere. If one adds the possibility of controlling the surface state, the radiative properties measurements may increase significantly the knowledge of optical and electrical properties of materials.

In this paper the radiative properties of two pure elements, copper and nickel, are widely studied. These metals were chosen since, at this moment, there is only a partial experimental knowledge of their radiative properties in the mid-infrared range. Nevertheless, both metals are employed in applications in which the knowledge of their radiative properties is very important. Copper is, for example, largely used for electrical wires, heat sinks and heat exchangers. Nickel is used to provide hardwearing decorative and engineering coatings, and plays an important role in alloys developed for a great number of specific and general purposes. Recently, both metals have been studied due to their effectiveness for decreasing the emissivity of coatings and as diffusion barrier layers between low-emissivity films and substrates [7,8]. Anyway, one can find some experimental data in the literature about the optical and electrical properties in the infrared spectral range of both metals. The total hemispherical emissivity of copper was measured at different temperatures [9–12]. Its normal spectral emissivity was studied at its melting point [13–15], above the melting point [16] and at 400 °C [17], but no directional spectral emissivity for copper can be found. In the case of nickel, its total normal emissivity was measured for different surface states [18], whereas its total hemi- spherical emissivity was obtained in the 77–737 °C range using calorimetric techniques [19]. The normal spectral emissivity was measured at 400 °C [17] using a radiometer, and a few disperse data were calculated from optical constants [20]. Additional spectral emissivity values were obtained between 1167 and 1332 °C for a high-purity nickel sample [21] and at its melting point with a pulse-heating technique [22]. The radiance temperature of nickel at its melting point [23] and the optical properties of liquid nickel were also measured [24]. However, a systematic study of its directional spectral emissivity as a function of the temperature has not been found. On the other side, emissivity values could be obtained using experimental optic constants values. However, for both cop- per and nickel, only experimental data at room temperature are available in the literature [3,25].

In this paper, the copper and nickel emissivity dependence on temperature, wavelength and emission angle is studied in the mid-infrared range using a high accurate radiometer. The results obtained in this study are compared to the few experimental measures available in the literature for the same temperature range. In addition, using the electromagnetic theory, experimental data are com- pared to literature data from optical and electrical measurements.

2. Experimental

High purity (>99.9%) samples of electrolytic copper and nickel have been used. Films of these metals with a thickness greater than 36 μm were deposited on 60 mm diameter and 3 mm thick Armco iron discs. The samples roughness was measured using a conventional rugosimeter and the obtained values are showed in Table 1. It must be noticed that the surface roughness has no significant effect on the infrared emissivity in some coatings [26], but it has noticeable effect in metals in the near infrared [27].

The emissivity measurements were carried out using a highly accurate radiometer (HAIRL) [28], which allows an accurate detection of the signal as well as its fast processing. A standard black- body is used for the radiometer calibration and an FT-IR spectrometer is used as a signal detector. A diaphragm within the spectrometer adjusts the sample area viewed by the detector. On the sample selected area a good homogeneity of temperature is ensured. The sample holder allows directional measurements and the sample chamber ensures a controlled atmosphere (vacuum, inert, oxidizing …). The sample temperature is measured by means of two K-type thermocouples spot-welded on the sample surface out of the area viewed by the detector. The sample surface is cleaned in an acetone ultrasonic bath before placing it in the sample holder.

Table 1

Summary of the samples surface roughness: roughness average ($R_a$), average maximum height ($R_z$), and maximum height of the profile ($R_t$).

| Sample | $R_a$ (μm) | $R_z$ (μm) | $R_t$ (μm) |
|---|---|---|---|
| Cu | 1.22 | 7.12 | 8.02 |
| Ni | 0.38 | 2.15 | 2.89 |

The HAIRL calibration is carried out using the modified two-temperature method that involves the measurement of blackbody interferograms at low and high temperature [29]. The sample emissivity is obtained using the blacksur method [30], which includes the measurement of the sample interferogram and the sample and its surrounding temperatures together with the calibration data. These choices are supported by the accuracy analysis of several methods for direct emissivity measurement [30] and the analysis of the emissivity error due to the calibration process as well as the short and long-term temporal stability of the calibration processes [29]. Additionally, the combined standard uncertainty of this direct emissivity device was previously obtained from the analysis of all uncertainty sources [31]. For the measurements presented in this paper the maximum combined standard uncertainty varies be- tween 1% and 10% depending on wavelength and temperature, its average value being around 4%. However, for low emissivity values at low temperatures and long wavelengths the uncertainty strongly increases, reaching values even higher than 20%.

All the measurements have been made following the same experimental procedure. Once a sample is introduced in the sample chamber a moderate vacuum is made inside. Then, before the sample is heated up to the measurement temperatures, a slightly reducing atmosphere ($N_2$ + 5% $H_2$) is introduced into the chamber in order to prevent the oxidation of the sample surface. The measurements are carried out during five successive heating thermal cycles between room temperature and 900 °C approximately. For each heating cycle the emissivity is measured at seven temperatures between 250 °C and the maximum temperature with temperature steps of around 100 °C. After the fifth heating cycle X- ray diffraction and electronic microscopy were used to check possible signs of oxidation on the sample surfaces.

3. Results and discussion

3.1. Normal spectral emissivity

Measurements of normal spectral emissivity of each sample have been carried out between 3 and 21 μm for each cycle. Fig. 1 shows the normal spectral emissivity at low and high temperature obtained for copper and nickel during the first and fifth thermal cycles. Both metals have low emissivity, but, in any case, the extremely low emissivity values of copper (Fig. 1(a)) are remarkable. According to the electromagnetic theory, the emissivity decreases as wavelength increases in both metals [1,2]. On the other side, the experimental results show a significant decrease in the emissivity values between the first to the fifth thermal cycle. A similar behaviour was previously seen in other metallic materials such as Armco iron and Ti–6Al–4V alloy [32,33]. This behaviour is associated to a surface stress relaxation process. The first heating cycles acts as an annealing process that relieves the surface stresses. This explains why the emissivity during the fifth thermal cycle is lower than during the first one and the difference in the emissivity value between the first and fifth thermal cycles in Fig. 1 is larger for the lowest temperature than for the highest. This is due to the measurement method used in this study, in which the temperature is stabilized during 20 min every 100 °C approximately in order to measure the emissivity. Thus, during the first cycle, at around 250 °C, the sample still retains all the surface stress, but, at around 850 °C, it has accumulated a significant stress relaxation produced in the previous temperature steps. Therefore, during the first heating cycle the relaxation kinetics has already begun before the last temperature step. This feature was extensively analysed for titanium alloys [33]. Other factors, such as oxidation or roughness, could be involved in the observed decrease of the emissivity with thermal cycling. However, the surface sample was analysed after the five heating cycles by means of X-ray diffraction and no signs of oxidation were found. In addition, the surface oxidation process usually produces an increase of the metal emissivity [34,35]. On the other hand, the emissivity values show a noticeable dependence on the roughness, but no substantial roughness changes are usually detected during the heating cycles [32,35].

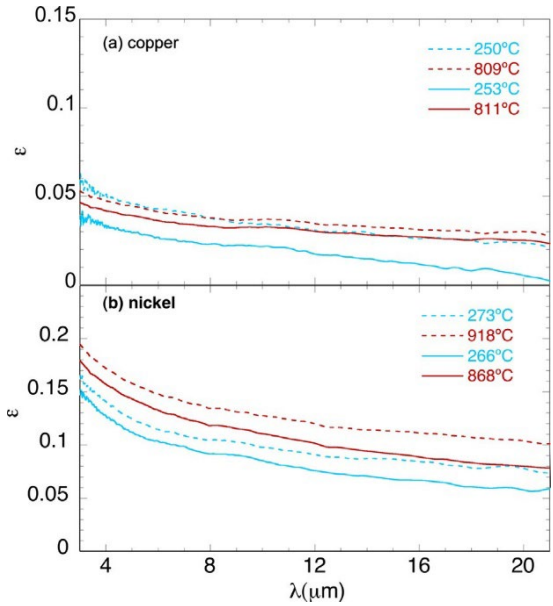

Fig. 1. Normal spectral emissivity of copper and nickel as a function of wavelength for two temperatures and two thermal cycles. Dashed and solid lines refer to the first and fifth heating cycles, respectively.

Once the surface stress has been relieved and the emissivity has reached a steady situation, one can analyse the emissivity dependence on other parameters. The copper and nickel emissivity spectra corresponding to the fifth heating cycle are plotted in Fig. 2 at different temperatures. Even though the average error value is around 4%, due to the low emissivity values, copper measurements at 253 °C show errors larger than 20% at λ > 18 μm. Regarding the emissivity dependence on the wavelength, nickel shows the normal progressive emissivity decrease predicted by the electromagnetic theory, but copper presents an irregular behaviour. In this case, a relative emissivity maximum appears between 8 and 12 μm (Fig. 2(a)). This relative maximum does not appear in the Armco iron [32] or nickel samples, so it can be ensured that it is an effect associated to copper. This type of behaviour is predicted for noble metals in mid and far infrared by a semi-classical theory of the optical properties of metals in the regime of the anomalous skin effect [36]. A metal is in this regime when the electron mean free path is comparable with the classical skin depth. The calculations carried out for copper show that the experimental results of this paper are in excellent qualitative agreement with the broad peak predicted by the anomalous skin effect theory in the mid- infrared [37].

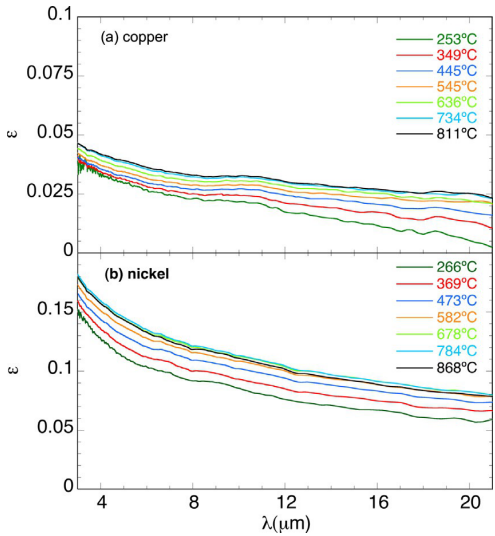

Fig. 2. Copper and nickel spectral emissivity for the fifth heating cycle at different temperatures.

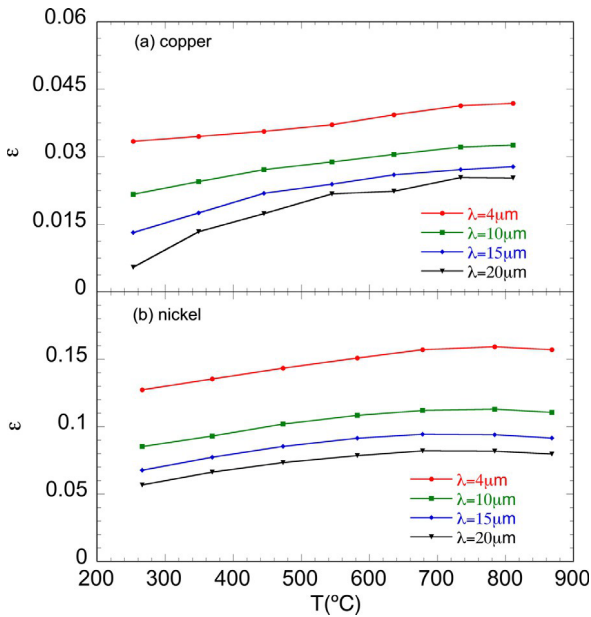

Fig. 3. Normal emissivity of copper and nickel obtained during the fifth heating cycle as a function of temperature for different wavelengths.

Considering the emissivity dependence on temperature, a clear increase can be observed in both copper and nickel (Fig. 2). This dependence is better visualized in Fig. 3, where the emissivity is plotted as a function of temperature at different wavelength. Usually, metals present a linear dependence with the temperature. However, it is remarkable that, in the case of nickel, the emissivity changes little and even slightly decreases when the temperature increases for T > 700 °C.

Fig. 4 compares the spectral emissivity values of this study with the literature emissivity data of pure copper and nickel. In these comparisons, it must be taken into account that some differences can easily appear due to experimental uncertainties, different

sur- face states and experimental conditions. Anyway, there is a good agreement in the comparison with the literature data at 400 °C [17] (Fig. 4(a) and (b)). In the case of copper, the discrepancies appearing at short wavelengths can be attributed to the fact that the sample used in Ref. [17] is a polished one. There are also some nickel literature data at 1167 °C [21]. If these data are compared to our experimental data at our maximum temperature 868 °C (Fig. 4(c)), there are important discrepancies due to the temperature differences. The discrepancies for λ >6 µm can be explained taking into account the nickel emissivity slightly decrease observed for T > 700 °C (Fig. 3(b)), whereas the divergences for λ <6 µm can be due to the detector accuracy or the beginning of an oxidation process in the emissivity measurements of Ref. [21]. It is known that the oxidation processes produce an emissivity increase at short wavelengths at their earlier stage. It is also important to notice that these comparisons show that the emissivity of copper and nickel bulk samples are similar to the ones of film samples.

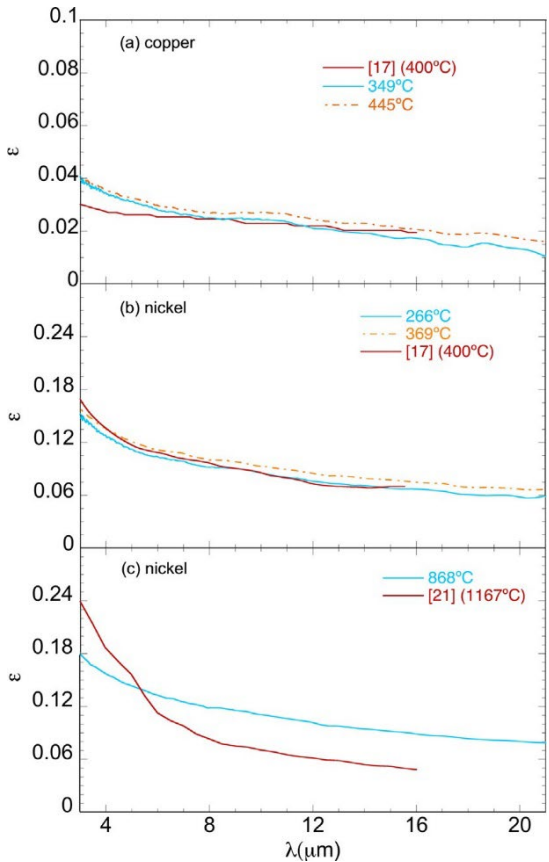

Fig. 4. Measured spectral emissivity of copper and nickel compared to the spectra of Refs. [17,21].

3.2. Directional spectral emissivity

The classical electromagnetic theory for polished surface metals predicts that the directional emissivity increases with increasing angle from the surface normal till near 90°, and it drops toward zero at this angle [1,2]. The directional spectral emissivity measurements were carried out for discrete angles between 10° and 80° during the fifth heating cycle. In Fig. 5 the directional emissivity of copper at λ =6 µm and nickel at

λ = 10 μm are displayed as a function of the emission angle for several temperatures. Similarly, the directional emissivity of nickel at T = 574 °C is displayed as a function of the emission angle for several wavelengths in Fig. 6. It can be observed in both figures that the emissivity shows the in- crease with the angle predicted by the electromagnetic theory for metallic samples. It is also observed in Fig. 6 that the emissivity curves for different wavelengths are closer at high emission angles. This way the emissivity spectra tend to present weak wavelength dependence at higher emission angles. This is in agreement with other experimental results for metallic samples [35]. Additionally, an irregular behaviour appears at high emission angles where the emissivity curve for λ = 20 μm crosses other curves. This means that at high emission angles the normal wavelength dependence changes and an irregular increase with the wavelength appears. This can also be observed in Ref. [35] and could be related to the roughness influence on the emissivity angular dependence.

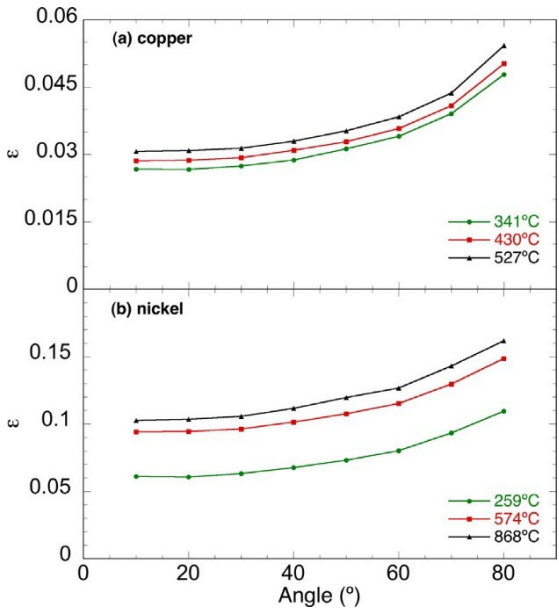

Fig. 5. Emissivity versus emission angle for three temperatures at λ =6 μm for copper and λ = 10 μm for nickel.

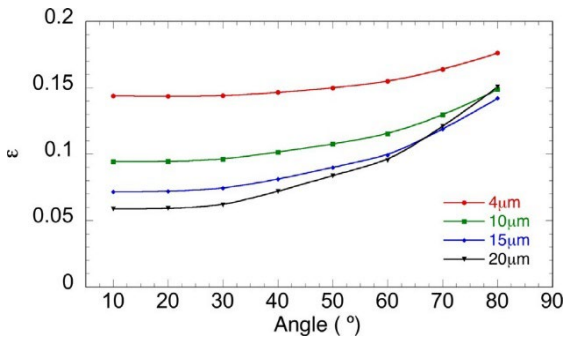

Fig. 6. Emissivity versus emission angle for nickel at 574 °C and four wavelengths.

3.3. Electric resistivity determination

The classical theory for an ideal interaction between the incident electromagnetic wave and the surface allows one to obtain useful relations between radiative, optical and electrical properties of metals. The variations of the measured property values from theoretical predictions are usually associated with the differences between the real materials and the ideal ones assumed by the theory. Additionally, the low accuracy of the experimental values of the constant optics for λ > 10 μm must be taken into account. In any case, the classical theory provides a number of useful tools. Thus, for perfectly flat metal surfaces, theory gives a simple relation between emissivity (ε) and electrical resistivity ($r_e$), the Hagen–Rubens equation [1,2]:

$$\varepsilon(\lambda) = 36.5 \cdot \sqrt{\frac{r_e}{\lambda}} - 464 \cdot \frac{r_e}{\lambda}, \tag{1}$$

where λ is the wavelength in μm and $r_e$ is the electrical resistivity in Ω-cm. This equation is a good qualitative approach, and sometimes also exhibits a good quantitative agreement for λ >5 μm wave- lengths. Eq. (1) enables us to use the available electrical resistivity data of copper and nickel [38,39] to obtain emissivity values. Analogously, the electrical resistivity can be obtained by fitting the experimental emissivity curves to the Hagen–Rubens equation. Fig. 7 shows this fitting for copper and nickel at 811 and 678 °C, respectively. The agreement is good in all the spectral range of emissivity measurements. If one does this fitting for all the temperatures, the electrical resistivity as a function of temperature is obtained. Fig. 8 shows the electrical resistivity of copper and nickel calculated using this method during the fourth and fifth thermal cycle together with the literature resistivity data [38,39]. There is a good qualitative agreement between direct measured resistivity values of copper and those calculated using Eq. (1). This shows that the primary contribution to the thermal radiation phenomena in copper is the intraband transition of electrons [40]. This feature is in agreement with the possibility of an anomalous skin effect regime in copper above room temperature, where the theoretical calculation of skin depth and mean free path have similar values [37]. For nickel a clear quantitative discrepancy was found between the literature electrical resistivity data and the values obtained using Eq. (1). This result agrees with the fact that nickel does not follow the free electron model for λ < 20 μm [5,6] and therefore neither the Hagen–Rubens equation. Finally, it is interesting to remark that even though the calculation of the electrical resistivity by using the emissivity measurements has not a practical use, it allows one to check the Hagen–Rubens approximation. For metals that fulfil this equation it will be useful to determine the infrared emissivity by using electrical resistivity measurements, which are easier to perform.

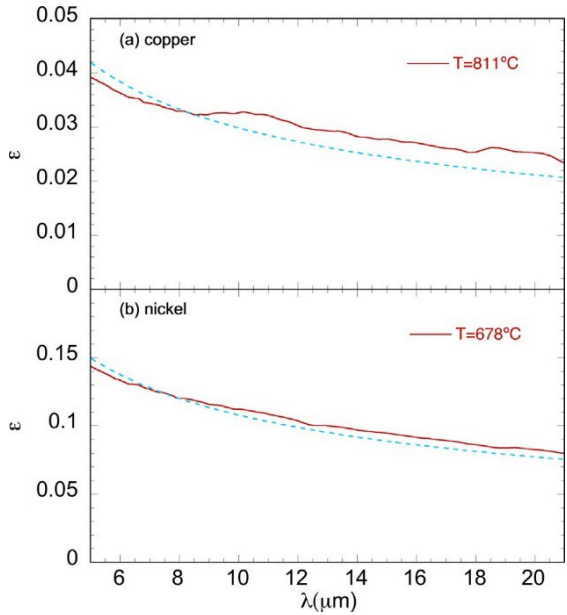

Fig. 7. Copper and nickel emissivity at 811 and 678 °C, respectively. Solid lines are the experimental data and dashed lines the emissivity values using the Hagen– Rubens relation.

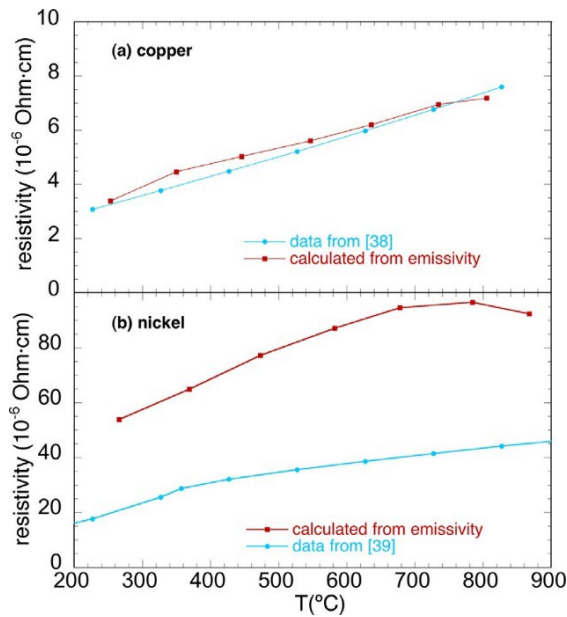

Fig. 8. Electrical resistivity dependence on temperature. Comparison between the values calculated from emissivity measurements using the Hagen–Rubens relation and the electrical resistivity data in Refs. [38,39].

4. Conclusions

Measurements of emissivity properties for copper and nickel, which are relevant to the scientific, energy and industrial processes, have been carried out. It was found that the spectral emissivity of these metals, in agreement with the electromagnetic theory, decreases as wavelength increases. However, an anomalous behaviour is found for copper. For this metal the spectral emissivity shows a broad plateau around λ = 10 μm. This broad plateau is predicted, for the noble metals in the mid and far-infrared, by a semi-classical theory of the optical properties of metals in the regime of the anomalous skin effect. In addition, the emissivity for both metals increases with temperature except for nickel above 700 °C. In this case, a different situation occurs and it is found that the spectral emissivity decreases when the temperature increases, which has not been observed for other pure metals. Additional work is needed to clarify this effect. Both metals show the usual emissivity increase with the emission angle, as predicted by the electromagnetic theory. However, the angular emissivity behaviour of nickel for λ > 20 μm differs at all analysed temperatures from those for lower wavelengths. For metals, the electromagnetic theory gives a simple relation (the Hagen–Rubens equation) between the emissivity (ε) and the electrical resistivity ($r_e$), which permits to use electrical measurements to obtain emissivities and vice versa. Comparison of the electrical resistivity results for the two methods, Hagen–Rubens equation applied to radiometric measurements and experimental resistivity data, lead to the conclusion that this prediction is fairly good for copper but is poor for nickel.

Acknowledgements

This research was partially supported by the program ETORTEK of the Consejería de Industria of the Gobierno Vasco in collaboration with the CIC-Energigune Research Center. L. González-Fernández acknowledges the Basque Government and Industria de Turbo Populsores S.A. their support through a Ph.D. fellowship.

References

[1]     R. Siegel, J. Howell, Thermal Heat Transfer, fourth ed., Taylor & Francis, Washington, 2002.

[2]     M.F. Modest, Radiative Heat Transfer, second ed., Academic Press, San Diego, 2003.

[3]     E.D. Palik (Ed.), Handbook of Optical Constants of Solids, Academic Press, San Diego, 1985.

[4]     M.A. Havstad, S.A. Self, Sensitivities of measurement methods for the thermal radiative properties and optical constants of metals in the spectral range 0.4 to 10 μm, Int. J. Thermophys. 14 (1993) 1077.

[5]     M.A. Ordal, R.J. Bell, R.W. Alexander Jr., L.L. Long, M.R. Querry, Optical properties of fourteen metals in the infrared and far infrared: Al Co, Cu, Au, Fe, Pb, Mo, Ni, Pd, Pt, Ag, Ti, V, and W, Appl. Opt. 24 (1985) 4493–4499.

[6]     A.D. Rakic, A.B. Djurisic, J.M. Elazar, M.L. Majewski, Optical properties of metallic fiμms for vertical-cavity optoelectronic devices, Appl. Opt. 37 (1998) 5271–5283.

[7]     H.J. Yu, G.Y. Xu, X.M. Shen, X.X. Yan, R. Huang, F.L. Li, Preparation of leafing Cu and its application in low infrared emissivity coatings, J. Alloys Compd. 484 (2009) 395–399.

[8]     X. Yan, G. Xu, Corrosion and mechanical properties of polyurethane/Al composite

coatings with low infrared emissivity, J. Alloys Compd. 491 (2010) 649–653.

[9] K.G. Ramanathan, S.H. Yen, E.A. Estalote, Total hemispherical emissivities of copper, aluminum, and silver, Appl. Opt. 16 (1977) 2810–2817.

[10] R. Smalley, A.J. Sievers, The total hemispherical emissivity of copper, J. Opt. Soc. Am. 68 (1978) 1516–1518.

[11] E.A. Estalote, K.G. Ramanathan, High-temperature emissivities of copper, aluminum, and silver, J. Opt. Soc. Am. 67 (1977) 32–38.

[12] E.A. Estalote, K.G. Ramanathan, Low-temperature emissivities of copper and aluminum, J. Opt. Soc. Am. 67 (1977) 39–44.

[13] H. Watanabe, M. Susa, K. Nagata, Near-Infrared spectral emissivity of Cu, Ag, and Au in the liquid and solid states at their melting points, Int. J. Thermophys. 24 (2003) 1105–1120.

[14] H. Watanabe, M. Susa, K. Nagata, Discontinuity in normal spectral emissivity of solid and liquid copper at the melting point, Metall. Mater. Trans. A 28 (1997) 2507–2513.

[15] C. Cagran, G. Pottlacher, Normal spectral emissivities of liquid copper, liquid gold and liquid silver at 684.5 nm, J. Non-Cryst. Solids 353 (2007) 3582–3586.

[16] K. Nagata, T. Nagane, M. Susa, Measurement of normal spectral emissivity of liquid copper, ISIJ Int. 37 (1997) 399–403.

[17] W. Bauer, H. Oertel, M. Rink, Spectral emissivities of bright and oxidized metals at high temperatures, in: 15th Symposium on Thermophysical Properties, Boulder, CO, USA, 2003.

[18] M. Misale, C. Pisoni, G. Tanda, Influence of surface finishing operation on total normal emittance of nickel and titanium, Int. J. Heat Technol. 6 (1988) 97–110.

[19] S.X. Cheng, Total hemispherical emissivities of cobalt and nickel in the range 350–1000 K, Exp. Therm. Fluid Sci. 2 (1989) 165–172.

[20] T. Makino, H. Kawasaki, Study of the radiative properties of heat resisting metals and alloys (1st report, optical constants and emissivities of nickel, cobalt and chromium), Bull. Jpn. Soc. Mech. Eng. 25 (1982) 804–811.

[21] G. Teodorescu, P.D. Jones, R.A. Overfeld, B. Guo, Normal emissivity of high-purity nickel at temperatures between 1440 and 1605 K, J. Phys. Chem. Solids 69 (2008) 133–138.

[22] K. Boboridis, A. Seifter, A.W. Obst, D. Basak, Radiance temperature and normal spectral emittance (in the wavelength range of 1.5 to 5 μm) of nickel at its melting point by a pulse-heating technique, Int. J. Thermophys. 28 (2007) 683–696.

[23] E. Kaschnitz, J.L. McClure, A. Cezairliyan, Radiance temperatures (in the wavelength range 530 to 1500 nm) of nickel at its melting point by a false- heating

technique, Int. J. Thermophys. 19 (1998) 1637–1646.

[24] S. Krishnan, P.C. Nordine, Optical properties of liquid nickel and iron, Phys. Rev. B 55 (1997) 8201–8206.

[25] F. Abelés (Ed.), Optical Properties and Electronic Structure of Metals and Alloys, North-Holland Publishing Co, Amsterdam, 1966.

[26] X.M. Shen, L.S. Li, X.R. Wu, Z.F. Gao, G.Y. Xu, Infrared emissivity of Sr doped lanthanum manganites in coating form, J. Alloys Compd. 509 (2011) 8116– 8119.

[27] C.-D. Wen, I. Mudawar, Emissivity characteristics of roughened aluminum alloy surfaces and assessment of multispectral radiation thermometry (MRT) emissivity models, Int. J. Heat Mass Transfer 47 (2004) 3591–3605.

[28] L. del Campo, R.B. Pérez-Sáez, X. Esquisabel, I. Fernández, M.J. Tello, New experimental device for infrared spectral direction emissivity measurements in a controlled environment, Rev. Sci. Instrum. 77 (2006) 113111.

[29] L. González-Fernández, R.B. Pérez-Sáez, L. del Campo, M.J. Tello, Analysis of calibration methods for direct emissivity measurements, Appl. Opt. 49 (2010) 2728– 2735.

[30] R.B. Pérez-Sáez, L. del Campo, M.J. Tello, Analysis of the accuracy of methods for the direct measurement of emissivity, Int. J. Thermophys. 29 (2008) 1141– 1155.

[31] L. del Campo, R.B. Pérez-Sáez, L. González-Fernández, M.J. Tello, Combined standard uncertainty in direct emissivity measurements, J. Appl. Phys. 107 (2010) 113510.

[32] L. del Campo, R.B. Pérez-Sáez, M.J. Tello, X. Esquisabel, I. Fernández, Armco iron normal spectral emissivity, Int. J. Thermophys. 27 (2006) 1160–1172.

[33] L. González-Fernández, E. Risueño, R.B. Pérez Sáez, M.J. Tello, Infrared normal spectral emissivity of Ti–6Al–4V alloy in the 500–1150 K temperature range, J. Alloys Compd. 541 (2012) 144–149.

[34] L. del Campo, R.B. Pérez-Sáez, M.J. Tello, Iron oxidation kinetics study by using infrared spectral emissivity measurements below 570 °C, Corros. Sci. 50 (2008) 194– 199.

[35] L. del Campo, R.B. Pérez-Sáez, L. González-Fernández, X. Esquisabel, I. Fernández, P. González-Martín, M.J. Tello, Emissivity measurements on aeronautical alloys, J. Alloys Compd. 489 (2010) 482–487.

[36] G.E.H. Reuter, E.H. Sondheimer, The theory of the anomalous skin effect in metals, Proc. R. Soc. A195 (1948) 336.

[37] T. Echániz, I. Setién-Fernández, R.B. Pérez-Sáez, M.J. Tello, Experimental verification of the anomalous skin effect in copper using emissivity measurements, Appl. Phys. Lett. 102 (2013) 244106.

[38] C.Y. Ho, M.W. Ackerman, K.Y. Wu, T.N. Havill, R.H. Bogaard, R.A. Matula, S.G. Oh, H.M. James, Electrical resistivity of ten selected binary alloy system, J. Phys. Chem. Ref. Data 12 (1983).

[39] F.C. Schewerer, L.J. Cuddy, Spin-disorder scattering in iron- and nickel-base alloys, Phys. Rev. B 2 (1970) 1575–1587.

[40] N.W. Ashcroft, N.D. Mermin, Solid State Physics, Saunders Collage Publishing, 1976.